\documentclass[
groupedaddress,
twocolumn, 
notitlepage,
bibnotes,
aps, prl, 10pt]{revtex4-2}

\usepackage{graphicx}
\usepackage{physics}
\usepackage{amsmath}
\usepackage{amssymb}
\usepackage{cancel}
\usepackage{empheq}
\usepackage{xcolor}
\usepackage[colorlinks,citecolor=blue,urlcolor=blue,linkcolor=blue,hypertexnames=true]{hyperref}
\usepackage{tabularx}
\newcolumntype{A}{>{\centering\arraybackslash}X}
\usepackage{array, makecell}

\newcommand{\etal}{\textit{et al.}~}

\newmuskip\pFqmuskip
\newcommand*\pFq[6][8]{%
  \begingroup % only local assignments
  \pFqmuskip=#1mu\relax
  \mathchardef\normalcomma=\mathcode`,
  \mathcode`\,=\string"8000
  \begingroup\lccode`\~=`\,
  \lowercase{\endgroup\let~}\pFqcomma
  {}_{#2}F_{#3}{\left[\genfrac..{0pt}{}{#4}{#5}~\middle|~#6\right]}%
  \endgroup
}
\newcommand{\pFqcomma}{{\normalcomma}\mskip\pFqmuskip}

\newcommand{\aell}{\abs{\ell}}

\newcommand{\qeff}{{q_\textrm{eff}}}

\begin{document}

\title{\textbf{Controlled generation of high-harmonic spatiotemporal optical vortices}}
\author{Titouan Gadeyne}
\email{titouan.gadeyne@cea.fr}
\author{Vartika Vishnoi}
\author{Thierry Ruchon}
\email{thierry.ruchon@cea.fr}
\affiliation{%
Universit\'e Paris-Saclay, CEA, LIDYL, 91191 Gif-sur-Yvette, France
}

\begin{abstract}
Spatiotemporal optical vortex (STOV) pulses are puzzling states of light, which see their transverse orbital angular momentum being debated in spite of possessing a topological charge. Thus far studied at conventional optical wavelengths, we herein demonstrate their upconversion to extreme ultraviolet (XUV) frequencies via high-harmonic generation (HHG). Circularly-symmetric infrared STOV pulses of topological charges $\ell=\pm1$ and $\pm2$ from a bespoke pulse shaper are focused in argon to drive harmonics, each order $q$ forming a spatiospectral ring in the far-field. Dependence of their radii on $q$ and $\ell$ indicates XUV vortices of charge $\ell_q=q\cross\ell$, and sensitivity of these modes to the focus position further establishes control over their production. The availability of XUV STOVs unlocks investigations and applications of their elusive angular momentum in the context of photoionization and other ultrafast spectroscopies.
\end{abstract}

\date{\today}

\maketitle

The realization that electromagnetic fields contain angular momentum (AM) has time and again played a decisive role in our understanding of light and its interactions with matter \cite{andrews_2012_angular, bliokh_2015_transverse}. Initial studies centered on \emph{spin} angular momentum (SAM) exerted profound influences on optics at the fundamental level, and enabled dichroic spectroscopies that now exploit both longitudinal and transverse components of the SAM vector \cite{
berova_2000_circular,
ranjbar_2009_circular, 
aiello_2015_transverse}. Renewed interest in the AM of light sparked in the 1990s when optical vortex beams were found to carry \emph{orbital} angular momentum (OAM) \cite{allen_1992_orbital}: their transverse amplitude distribution features a phase singularity, endowing them with an OAM vector parallel to their propagation axis. These have become a cornerstone of modern photonics, with physical consequences of their structure unveiled in optical manipulation, quantum optics, atomic spectroscopy, and more \cite{shen_2019_optical}. The picture of OAM-carrying light was recently completed by spatiotemporal optical vortex (STOV) pulses \cite{hancock_2019_free, chong_2020_generation}: wavepackets with a phase vortex embedded in the \emph{space-time} domain. These gave the first example of light transporting an intrinsic OAM vector \emph{perpendicular} to its propagation: from a theoretical standpoint, this one has proven to be highly more elusive than that of spatial vortices, and remains actively discussed \cite{
hancock_2021_mode,
bliokh_2023_orbital,
porras_2024_clarification, 
gadeyne_2025_energy}. Experimental implementations of STOV pulses hold the promise to deepen our understanding of complex light-matter interactions, by unlocking previously inaccessible configurations.

If shaping and characterization of complex pulses are now well-developed at optical wavelengths, they remain challenging at photon energies in the extreme-ultraviolet (XUV) and X-ray domains. These support high-resolution, element-specific and attosecond-time-resolved spectroscopies, that already benefit from harnessing the SAM of light \cite{laan_2014_x, vaz_2025_x}. In this context, high-harmonic generation (HHG) emerged as an attractive route: this highly coherent nonlinear frequency-conversion process enables the phase structure of a driving laser pulse to be transferred onto its high-order harmonics. Using this approach, spatial XUV vortices were demonstrated a decade ago \cite{gariepy_2014_creating, geneaux_2016_synthesis}, and are showing potential for high-resolution imaging \cite{eschen_2022_material} and probing of magnetic textures \cite{fanciulli_2022_observation}
or chiral molecules \cite{rouxel_2022_hard}. Extending STOVs to XUV wavelengths was soon recognized as an objective for future investigations \cite{shen_2023_roadmap}, and simulations predicted that driving HHG with a STOV of topological charge (TC) $\ell$ would generate high-harmonic STOV pulses, with the $q^\textrm{th}$ harmonic order exhibiting a TC $\ell_q = q \cross \ell$ \cite{fang_2021_controlling}. 
Closely related experimental results were since reported. First, doubling of the TC was verified in STOV-driven second-harmonic generation ($q=2$) \cite{gui_2021_second, hancock_2021_second, gao_2023_spatiotemporal, liang_2025_cascaded}, demonstrating their conversion by a low-order, perturbative process. More recently, a mainly theoretical study of STOV-driven HHG was complemented by experimental observations for which imperfect, non-circular driving STOVs were used \cite{martinhernandez_2025_extreme}. While in fair agreement with the corresponding simulations, the reported harmonic modes failed to exhibit the full ring-like, vortex structure expected from ideal STOV-driven HHG.

In this Letter, we demonstrate the controlled upconversion of a STOV pulse through high-harmonic generation. Building upon these previous studies, we designed a pulse-shaping scheme outputting energetic, nearly circular STOVs from femtosecond laser pulses. The ring-like spatiospectral structure of their harmonics and its variation with $q$ and $\ell$ confirms the embedding of a $\ell_q = q \cross \ell$ phase vortex within the XUV pulses. We report how this harmonic mode evolves as a function of the position of the nonlinear medium, reflecting the propagation-induced reshaping inherent to the driving STOV pulse in excellent agreement with simulations.

\begin{figure}[t!]
    \centering
    \includegraphics[width=\linewidth]{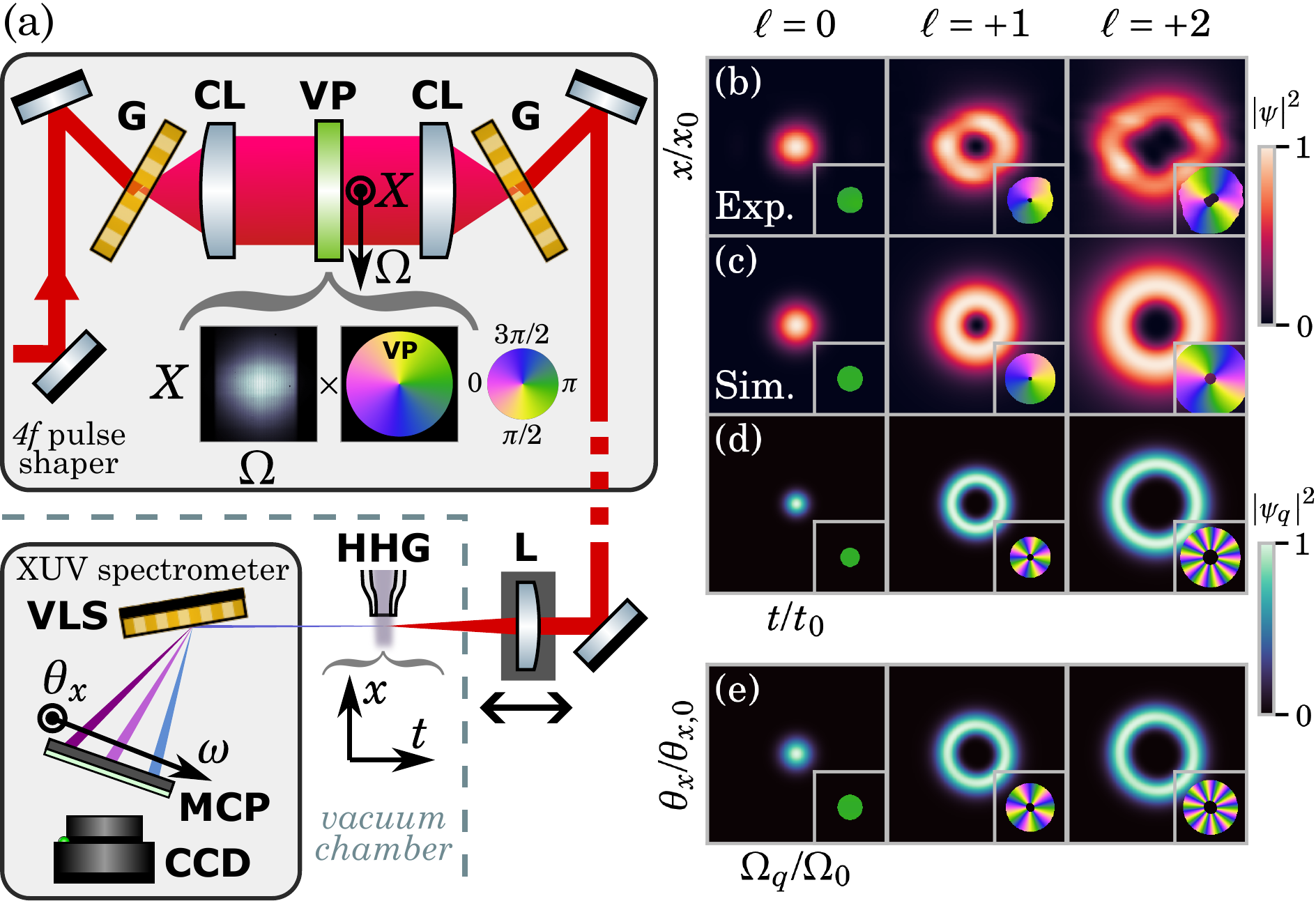}
    \caption{
    (a) Experimental setup for STOV pulse shaping and spatiospectral imaging of high-harmonics.
    (b) Experimental and (c) theoretical space-time pulse envelopes $\psi(x,t)$ at the focal plane $z=f$, for $\ell = 0$, $+1$ and $+2$. The phase $\arg[\psi]$ is shown in inset on a cyclic color scale, and blanked where intensity drops below $10$\% of its peak value. Here theory assumes $f=1$~m as in our pulse measurement setup.
    (d) Theoretical spatiotemporal envelope $\psi_q(x,t)$ of harmonic $q=5$ in the nonlinear medium, and (e) spatiospectral envelope $\widetilde{\psi}_q(\theta_x, \Omega_q)$ after propagation to the far-field. A low $q$ is deliberately used for the TC $\ell_q = q \cross \ell$ to be visible. Phase is blanked at intensities $<1$\%. Here $f=15$~cm is assumed, as in our HHG stage.
    For (b-e), the ranges shown are $\pm 3x_0$ for $x$, $\pm 3t_0$ for $t$, $\pm 3\theta_{x,0}/\sqrt{\qeff}$ for $\theta_x$, $\pm 3\Omega_0q/\sqrt{\qeff}$ for $\Omega_q$; in (b), $x_0$ and $t_0$ were extracted from the $\ell=0$ data.
    }
    \label{fig1}
\end{figure}

The experiment was carried out on the ATTOLab platform at CEA Saclay. We used an Yb-based laser source delivering $2$~mJ, near Fourier-limited pulses ($1/e^2$ half-duration $t_0 = 267$~fs), around a wavelength $\lambda_0 = 1033.6$~nm ($k_0 = 2\pi/\lambda_0$, $\omega_0 = ck_0$). We followed the popular approach to producing STOV pulses using a vortex plate (VP) inserted into a $4f$ pulse shaper \cite{hancock_2019_free, gui_2022_single}, as sketched in \autoref{fig1} (a). A first grating (G, groove density $N_s = 1000$/mm) disperses all frequencies horizontally, and a cylindrical lens (CL, $f_s = 40$~cm) focuses them at different positions in the Fourier plane of the shaper. The vertical ($X$) dependence of the input field is unaffected, and at this plane, a spatiospectral $(X, \Omega = \omega-\omega_0)$ image of the incident field is realized. The experimental intensity distribution there is inset in \autoref{fig1} (a) and can be modeled by a Gaussian of $1/e^2$ radii $X_0 \approx 2.1$~mm and $\Omega_0 \approx 2/t_0$%= 7.5 \cdot 10^{-3}$~rad/fs
. A VP with TC $\ell$ is introduced here to imprint an azimuthal phase in the $(\Omega/\Omega_0, X/X_0) 
= (R \cos\Theta, R \sin\Theta)$ domain, such that the spatiospectral field reads
\footnote{As is customary in studies on STOVs, the $y$-dependence of the field is ignored throughout, assumed to be an unstructured, separable envelope.}
\begin{align}
    \widetilde{\psi}
    (X, \Omega)
    =
    e^{-R^2}
    e^{-i \ell \Theta}.
    \label{eq:psi_shaper}
\end{align}
We found empirically that a $6$-mm-diameter iris centered on the VP in the Fourier plane helped clean the space-time mode of the output STOVs, still letting through $>90$\% of the incident energy. This likely removed slight nonlinear deteriorations that arise when propagating an intense few-mm beam over several meters. Frequencies are recombined into a collimated beam by a second identical set of cylindrical lens and grating. At this stage, the resulting pulse is not yet a STOV: it forms after a spatial Fourier transform (FT), performed by a subsequent spherical lens (L, $f=15$~cm). Neglecting propagation between the Fourier plane and focusing lens, the space-time field $\psi(x, t) e^{-i \omega_0 t}$ at a distance $z$ beyond the lens reads
\begin{align}
\begin{split}
    \psi(x, t ; z)
    \propto
    \!\int\!\!\!\!\int\! \dd \Omega \dd X
    e^{-i\Omega t+ik_0[\frac{(X-x)^2}{2z} - \frac{X^2}{2f}]}
    \widetilde{\psi}(X, \Omega)
    .
\end{split}
\label{eq:psi_focus}
\end{align}
At $z=f$, this becomes proportional to a 2D FT over $(X, \Omega)$ of \eqref{eq:psi_shaper}, the result of which reads
\begin{align}
    \psi(x, t ; z=f)
    \propto
    r^\abs{\ell}
    e^{-r^2}
    e^{-i \ell \theta}
    \cdot 
    \pFq{1}{1}{\abs{\ell}/2}{\abs{\ell}+1}{r^2}
\label{eq:psi_focus_HyGG}
\end{align}
where we have introduced space-time polar coordinates at focus, 
$(t/t_0, x/x_0) 
=(r \cos\theta, r \sin\theta)$. Equation \eqref{eq:psi_focus_HyGG} describes a STOV pulse, with a phase vortex of charge $\ell$ and circularly-symmetric intensity distribution in a space-time plane. When using a VP (which only shapes the \emph{phase} of the incident field, not its \emph{amplitude}) the radial envelope obtained after a 2D FT does not exhibit the well-known Laguerre-Gaussian (LG) form: it involves an additional factor expressed above using the hypergeometric function $\pFq{1}{1}{a}{b}{z}$. Radial profiles of such modes decay slower than LG modes at large $r$, and their peak radius grows faster with $\ell$ \cite{karimi_2007_hypergeometric, bekshaev_2009_structure}. 
This expression proves more accurate than a LG approximation for estimating the pulse energies necessary for driving HHG, which typically requires reaching intensities $\gtrsim 5 \cdot 10^{13}$~W/cm$^2$. For our parameters, the required energy in a Gaussian ($\ell=0$) pulse would be $W \gtrsim 100$~µJ. In a STOV pulse however, intensity spreads over a large ring, leading to a much lower peak intensity for the same total energy: from \eqref{eq:psi_focus_HyGG}, we found that for $\aell = 1$ it is only $20$\% of that of the $\ell=0$ mode, while for $\aell=2$ it is a mere $8$\%. Correspondingly, the minimum energy rises to $500$ and $1250$~µJ in these two cases. Here, we achieved pulse shaping with $\approx 85$\% transmission, thus requiring a $\sim 1.5$ mJ input, within the range of our laser. These strict requirements are one reason for preferring $4f$ shaping and glass VPs over single-element STOV-shaping optics (\textit{e.g.} nanogratings \cite{huo_2024_observation}, or liquid crystal slabs \cite{alonso_2026_agile}), which may not yet offer the same conversion efficiency and resistance to high fluence.

To characterize infrared pulses, we implemented spatially-resolved Fourier-transform spectral interferometry (FTSI), following its demonstration as a convenient STOV imaging method \cite{gui_2022_single}; our FTSI setup will be described elsewhere. To fit it between the lens and focus, we imaged our pulses using a $1$-m focal length instead of the $15$-cm one used for HHG -- paraxially, this has no effect on the pulse shape apart from a rescaling of $x_0$ and a different wavefront curvature. The intensity and phase of our pulses are compared to the Gaussian theory \eqref{eq:psi_shaper}--\eqref{eq:psi_focus_HyGG} in \autoref{fig1} (b) and (c): the space-time phase vortex is clear, and good agreement is found in the ring size as $\ell$ increases. Most importantly, our STOVs display good circular symmetry: to achieve this, we selected the parameters of our pulse shaper ($N_s$ and $f_s$) to obtain a frequency spread $1/\alpha \approx 3.78 \cdot 10^{-3}$~(rad/fs)/mm in the Fourier plane \cite{monmayrant_2010_newcomer’s}, to match the extents of the spatial and spectral widths onto the vortex plate. This can be quantified by the \emph{inhomogeneity parameter} $\eta = X_0 / \alpha\Omega_0
\approx 1.07$, close to unity. This feature is crucial, as \cite{martinhernandez_2025_extreme} demonstrated that a large inhomogeneity ($\eta \approx 1.8$ for the reported experiments) prevents generating proper ring-like infrared and high-harmonic STOVs.

We simulate the harmonic field with a simple analytical model, validated across a variety of experiments involving space- or time-structured driving fields 
\cite{catoire_2016_complex, 
rego_2016_nonperturbative,
rego_2017_ultrashort,
chang_2021_high,
vimal_2023_photon, luttmann_2023_nonperturbative}. The phase of the $q^\textrm{th}$ harmonic field is $q \arg{\psi}$ (the high-harmonic dipole phase is ignored), while its amplitude scales as a power law $\propto \abs{\psi}^\qeff$; in practice we use $\qeff \approx 3.5$ for all orders based on numerical calculations \cite{suppmat}. We consider the nonlinear medium thin enough to disregard effects of longitudinal phase-matching (``thin-slab'' approximation \cite{rego_2016_nonperturbative}), as if HHG occurred at a single plane $z$ where the harmonic field $\psi_q e^{-i q \omega_0 t}$ thus reads
\begin{align}
    \psi_q(x, t; z) = \abs{\psi(x, t; z)}^\qeff e^{i q \arg[\psi(x, t; z)]}
    \label{eq:psiq_focus}
\end{align}
As harmonics cannot be diagnosed \textit{in situ} inside the generation medium, we observe them with an XUV spectrometer after diffraction to the far-field. Introducing the divergence angle $\theta_x$ and $\Omega_q = \omega - q \omega_0$, the spatiospectral envelope is obtained using the Fraunhofer integral,
\begin{align}
    \widetilde{\psi}_q(\theta_x, \Omega_q) 
    \propto
    \!\int\!\!\!\!\int\! \dd t\,\dd x
    \;
    e^{i \Omega_q t}
    e^{-i x k_x}
    \,
    \psi_q(x, t)
    \label{eq:psiq_farfield}
\end{align}
where $k_x = q k_0 \theta_x = 2 q \theta_x/\theta_{x,0}x_0$. The transform of the field given by \eqref{eq:psi_focus_HyGG} and \eqref{eq:psiq_focus} cannot be performed analytically. However, it can be shown \cite{chong_2020_generation} that any space-time vortex mode of the form $\psi_q \sim F(r) e^{-i \ell_q \theta}$ will turn into another vortex mode $\widetilde{\psi}_q \sim \widetilde{F}(\varrho) e^{+i \ell_q \vartheta}$ in the far-field spatiospectral domain with polar coordinates $(\Omega_q/\Omega_0, q \theta_x/\theta_{x,0}) 
= (\varrho \cos\vartheta, \varrho \sin\vartheta)$. 
\autoref{fig1} (d) and (e) show the theoretical harmonic mode in the spatiotemporal near-field \eqref{eq:psiq_focus} and spatiospectral far-field \eqref{eq:psiq_farfield}. Strikingly, STOV-driven harmonics should form a ring in the spatiospectral domain, with an on-axis minimum where conventional ($\ell=0$) HHG emission is usually observed.

High-order harmonics were generated inside a vacuum chamber, by focusing the shaped pulse into an argon gas jet. To minimize phase-matching effects, we used a thin jet ($100$-µm-diameter nozzle) with backing pressure from 2 to 5 bars. The XUV spectrometer consisted of a variable line spacing (VLS) grating (groove density $2400$/mm) followed by a set of microchannel plates (MCPs) and phosphor screen imaged by a CCD camera. The VLS at once disperses and focuses the harmonic frequencies horizontally, producing a spatiospectral image of the harmonic field onto the MCPs.
\begin{figure}[t!]
    \centering
    \includegraphics[width=\linewidth]{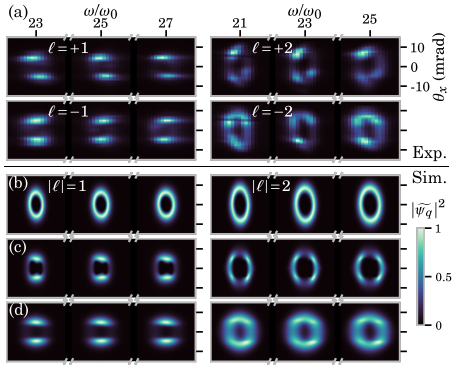}
    \caption{
    (a) Experimental far-field harmonic spatiospectra, driven by STOV pulses of TCs $\ell=\pm 1$ and $\pm 2$. Three successive harmonic orders $q$ are shown, with the axis cut beyond $\pm 0.1\omega_0$ around each order (the spectrum between orders is empty). All modes are normalized to their peak value.
    (b-d) Simulated harmonic modes \eqref{eq:psiq_farfield} for the $\abs{\ell} = 1$ and $2$ drivers (intensity distributions are identical for the two signs). (b) assumes the ideal Gaussian spatiospectrum \eqref{eq:psi_shaper} on the vortex plate; (c) uses our experimental spectrum. (d) is panel (c) convolved with a Gaussian kernel to model our experimental spectral resolution.
    }
    \label{fig2}
\end{figure}
%% Figure 2
In \autoref{fig2} (a), we report excerpts of high-harmonic spectra driven by STOV pulses of TCs $\ell=\pm1$ and $\pm2$. The intensity of every harmonic order distributes along a ring around a central minimum in the spatiospectral domain, its diameter expanding when increasing $\aell$. Harmonics driven by $+\ell$ and $-\ell$ STOVs are almost identical: this confirms that the gas jet was positioned near the focus, where STOV pulses were round and thus had the same intensity envelope for both signs of $\ell$. 

For $\aell = 2$, the ring shape of the harmonics is particularly convincing. For $\aell = 1$, the left/right arcs of the ring were always much fainter than the top/bottom lobes. Although a central minimum can still be seen, these appear different from ideal Gaussian STOV harmonics, simulated in \autoref{fig2} (b). Considering the agreement of our driving pulses with theory (\autoref{fig1}), we understood that this asymmetry was instead contributed by two specificities of our setup. First, our laser spectrum was not exactly Gaussian \cite{suppmat}: repeating the simulation with this specific spectrum and its spread $1/\alpha$ onto the vortex plate, we obtain \autoref{fig2} (c), which still exhibits spatiospectral vortices but with a less symmetrical intensity ring. Second, the bandwidth of each order approached the resolution limit of our spectrometer. While vertical resolution was limited only by camera pixelation and/or grains of the phosphor screen, harmonics were focused by the VLS grating on the horizontal axis, along which resolution gets limited by their spot size. In \autoref{fig2} (d), we model this by convolving panel (c) with a Gaussian response function $e^{-(\omega/\bar{\omega})^2}$, with $\bar{\omega} = 0.05\, \omega_0$ estimated by comparing the spectral widths of our $\ell=0$ harmonics to theory. This instrumental effect finally reproduces our observations: the left/right arcs spread out, making the central minimum difficult to observe. Note that without this artifact, the mode in (c) forms an entire ring, around which the phase winds through $q \cross \ell$ cycles without interruption. For the wider modes obtained with $\aell = 2$, the ring structure remains very clear with this modeling as well.

\begin{figure}[t!]
    \centering
    \includegraphics[width=\linewidth]{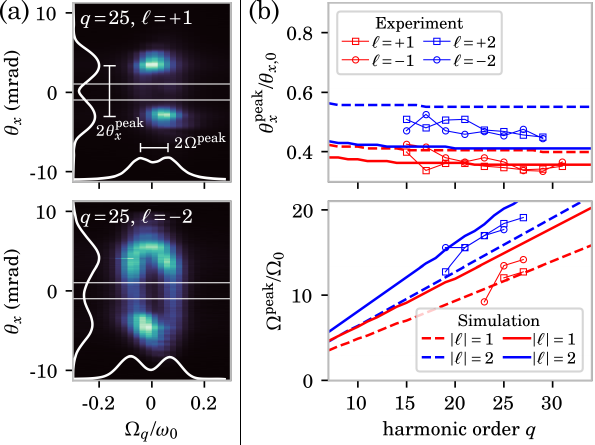}
    \caption{
    (a) Extraction of the spatial and spectral radii $\theta_x^\textrm{peak}$ and $\Omega^\textrm{peak}$, illustrated on two experimental harmonic rings. Spectral lineouts are obtained by integrating over $\theta_x$ between the two horizontal lines. The entire mode is integrated along $\Omega_q$ to obtain the spatial profiles. Radii are defined as half the peak-to-peak distance on these profiles. 
    (b) Evolution of the radii with $q$. Experimental data (symbols) is shown for four TCs of the driving STOV. Lines are radii obtained by the same procedure on numerical simulations, assuming an ideal Gaussian pulse (dashed) or the spectrum of our laser (full).
    }
    \label{fig3}
\end{figure}

%% Figure 3
The spatial extent of the rings in \autoref{fig2} appears roughly constant across all orders $q$. This observation is strictly analogous to one made 10~years ago by Géneaux \etal for HHG driven by \emph{spatial} vortex beams \cite{geneaux_2016_synthesis}. There, harmonic orders were also found to have the same diameter, in ($\theta_x, \theta_y$) space. This was explained as follows: assuming $\qeff$ to be independent of $q$, all orders share the same ring diameter at focus. However, they do not have the same transverse phase gradients, due to the TC $e^{iq\ell\theta}$ scaling with the harmonic order. As a result, their divergence should grow linearly with $q$. But in parallel, the wavelength decreases in inverse proportion to the harmonic order, reducing divergence by a factor $1/q$: the two effects cancel out to produce a $q$-independent ring size. In STOV-driven HHG, this reasoning holds for the \emph{spatial} radius. Along the \emph{spectral} dimension however, there is no diffraction hence no $1/q$ factor: the spectral radius should increase linearly with $q$. Observing these scalings is an indirect way of confirming the $\ell_q = q \cross \ell$ scaling of the TC embedded into the harmonic field. 

\autoref{fig3} (a) illustrates how we defined ring radii in practice. The peak intensity radius $\Omega^\textrm{peak}$ along the spectral dimension could be measured from a lineout taken near $\theta_x=0$. Given our low resolution along the spectral axis, it was better to quantify the spatial divergence via the peak intensity radius $\theta_x^\textrm{peak}$ of the profile obtained after integrating the full mode along $\Omega$. We plot experimental radii as a function of $q$ in \autoref{fig3} (b). They are independent of the sign of $\ell$, and increase with $\abs{\ell}$ along both dimensions. As expected from the assumption of a $\ell_q = q \cross \ell$ space-time phase vortex, the spatial radius is roughly independent of $q$, while the spectral radius increases linearly with it. We compare these results to the theory
\eqref{eq:psi_shaper}--\eqref{eq:psiq_farfield} using an ideal Gaussian spectrum (dashed lines), or that of our laser (full lines). Experiment and theory agree on the overall scalings with $q$, and absolute values of our experimental radii are within the range of these models. 

\begin{figure}[b!]
    \centering
    \includegraphics[width=\linewidth]{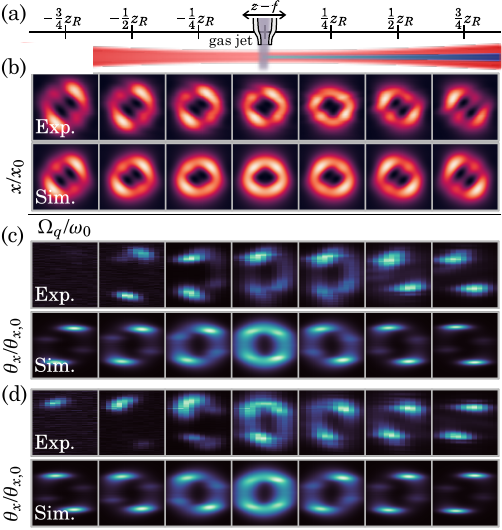}
    \caption{
    (a) Sketch of the position scan. The longitudinal offset $z-f$ between gas jet and laser focus could be varied over several $z_R$.
    (b) Experimental and theoretical space-time intensity of the $\ell=+2$ infrared STOV pulse at several positions $z-f$ (multiples of $z_R/4$).
    The simulation uses the experimental spectrum \cite{suppmat}. The plotted range is $\pm 3x_0$ for $x$, $\pm 3t_0$ for $t$.
    (c) (resp. (d)) Experimental and theoretical far-field spatiospectral intensity of harmonic $q=25$ for the corresponding jet positions, obtained with $\ell=+2$ (resp. $-2$) drivers. Simulations include instrumental broadening along the spectral axis as in \autoref{fig2} (d). The plotted range is $\pm \theta_{x,0}$ for $\theta_x$, $\pm 35\Omega_0$ for $\Omega_q$. All panels are normalized independently to their peak value to highlight modal evolutions; color bars are those of \autoref{fig1}.
    }
    \label{fig4}
\end{figure}

%% Figure 4
Another salient feature of STOV pulses is the evolution of their space-time envelope with propagation. This renders STOV-driven HHG uncommonly sensitive to placement of the nonlinear medium with respect to the focus. In conventional Gaussian-driven HHG this influences phase-matching \cite{balcou_1997_generalized}, whereas here the main observations will come from the reshaping of the driving pulse. The generating medium was thin compared to the Rayleigh length of the driver ($z_R \approx 1.9~$mm), hence translating the focusing lens over millimeters allowed probing the pulse at different stages of its reshaping, as sketched in \autoref{fig4} (a). Panel (b) compares experimental and theoretical envelopes of the $\ell=+2$ driving pulse at several distances from the focal plane, showing breakage of the circularly-symmetric ring into $\aell + 1$ lobes aligned along a diagonal -- the sign of this diagonal flips with that of $\ell$ \cite{porras_2023_propagation}. In (c), we compare the spatiospectral mode of harmonic $q=25$ obtained experimentally and numerically, for the $\ell=+2$ driver. As the jet is moved away, the ring-like mode breaks into two lobes arranged along a diagonal, the sign of which changes on either side of the focus in agreement with simulations: the $(\theta_x, \Omega_q)$ harmonic mode ``encodes'' the $(x, t)$ pulse in the jet, presented in (b). (d) features corresponding data for $\ell=-2$: all diagonals change sign, confirming that these observations are due to the phase structure introduced into the shaped pulse. Equivalent results for $\ell= \pm1$ are reported in the SM \cite{suppmat}.

%% CONCLUSION
To summarize, we have extended STOV pulses to XUV wavelengths via HHG, demonstrating that it can upconvert complex space-time phase topologies. Analysis of the harmonic modes confirms XUV vortices with the predicted high TC, $\ell_q = q \cross \ell$. We highlight careful monitoring of space-time wavepackets and their propagation as a vital step to harness them in nonlinear optics, which shall only become more relevant when moving to thicker media or non-paraxial geometries. Finally, we reached fair agreement with simulations throughout using a model implementing only the coherent and nonperturbative natures of HHG. We conclude that in our conditions, the dipole phase, macroscopic physics (phase-matching, reabsorption) and atom-dependent subtleties played minor roles and did not hinder the upconversion of the STOV structure. Accessible XUV STOVs could spark new experimental tests of light-matter interaction, wherein they might serve as probes with inherent sensitivity to space-time-coupled physics, excite electronic transitions with specific selection rules with regards to their transverse OAM, or photoionize electron wavepackets structured over space and time.
\\

\paragraph*{Acknowledgments}

This work was supported by state funding managed by the National Research Agency under the France 2030 investment plan bearing the reference ANR-22-EXLU-0002 and the moonshot project ``Tornado"; it received funding from the Agence Nationale
pour la Recherche (contract No. ANR HELIMAG ANR-21-
CE30-0037), and the Indo-French CEFIPRA Grant Project No.- (7104-I).

\paragraph*{Author contributions}

T.G. performed the experiments, analyzed the data, developed the analytical theory and wrote the first draft of the manuscript. V.V. helped conduct the experiments. T.R. conceptualized the work, provided guidance on all parts of the project and contributed to the manuscript.

\bibliographystyle{apsrev4-2}
\bibliography{allV2}

\end{document}